# Bulk electron spin polarization generated by the spin Hall current

V.L. Korenev

A. F. Ioffe Physical Technical Institute, St. Petersburg, 194021 Russia

*It is shown that the spin Hall current generates a non-equilibrium spin polarization in the interior of crystals with reduced symmetry in a way that is drastically different from the previously well-known "equilibrium" polarization during the spin relaxation process. The steady state spin polarization value does not depend on the strength of spin-orbit interaction offering possibility to generate relatively high spin polarization even in the case of weak spin-orbit coupling.*

*Corresponding author: korenev@orient.ioffe.ru*

Spin-orbit coupling of spin with orbital degrees of freedom creates spin current even in non-magnetic semiconductor. For example, the charge flow (both drift and diffusive) $\vec{j}$ of non-polarized electrons induces a spin Hall current [1, 2] even in high symmetry crystals: electrons with opposite spins deviate perpendicular to $\vec{j}$ in opposite directions. On the one hand, the spin Hall current leads to the anomalous Hall effect in the system of spin-polarized electrons with spin density $\vec{S}$, i.e. to the appearance of an additional transverse charge flow $\vec{j}_t \propto \vec{j} \times \vec{S}$ [3, 4]. On the other hand, it leads to the inverse effect [5], referred to as the spin Hall effect (SHE): the accumulation of non-equilibrium spin polarization near the sample edge (but not in the interior).

Semiconductors with reduced symmetry (for example, semiconductor quantum wells or strained GaAs) possess a linear in the momentum spin splitting of conduction band that brings about a number of new phenomena. Dyakonov and Kachorovski (DK) have shown [6], that the spin relaxation mechanism of Dyakonov-Perel [7] is enhanced and becomes anisotropic in quantum wells. It was observed [8] that electric current in the quantum well plane is accompanied by the appearance of the effective magnetic field $\vec{B}_{eff}$ bringing about the Larmor precession of electron spin. It was predicted [9] that (i) not only drift but also the diffusive flow of charge induces the average spin-orbit field $\vec{B}_{eff}$; (ii) the field $\vec{B}_{eff}$ is capable of rotating the light polarization plane, i.e. it may cause an electrically induced Faraday effect; (iii) the constant field $\vec{B}_{eff}$ will shift electron spin resonance (ESR) line; (iv) the $\vec{B}_{eff}$ alternating with ESR frequency will induce the ESR signal. References [10] and [11] predicted current-induced electron spin polarization. Spin relaxation process is very essential for this effect [12]. It has been interpreted [8, 11] as an "equilibrium" polarization: the spin polarization, $\vec{s}_T$, is proportional to $\mu_B \vec{B}_{eff}/T$ for Boltzmann statistics [8] or $\mu_B \vec{B}_{eff}/E_F$ for Fermi statistics [11] (T and $E_F$ are the temperature and Fermi energy). Kalevich-Korenev-Merkulov (KKM) proposed [13] a theory of the relationship between the non-equilibrium spin and spin current (spin flux) in a weak spin-orbit coupling regime in crystals with a linear in the momentum spin splitting. They found that spin current generates the non-equilibrium electron spin density. In turn, the spin polarization results in the appearance of spin current. The KKM theory presented a unified view of the DK spin relaxation and precession in averaged effective magnetic fields due to drift-diffusion motion of spin. However it did not take into account the spin Hall current.

Recent experiments [14, 15] have demonstrated the precession of the optically injected electron spin about the drift-diffusion induced spin-orbit field $\vec{B}_{eff}$ [8, 9] in strained bulk n-GaAs. One of the main challenges is to generate bulk electron spin polarization by the charge flow. Current-induced spin polarization resulted from the $\vec{B}_{eff}$ field has been observed in two-dimensional hole gas [16] in agreement with [10, 11, 12]. Similar effect was observed [17, 18] in n-type systems, too. However, according to Ref. [18], the origin of the current-induced electron polarization in strained bulk n-type GaAs crystals may come from the current-induced *generation* of spin rather than its relaxation to the "equilibrium" value $\rho_T$. It points to another mechanism of spin polarization of electrons. The experiments in n-type samples were carried out in weak spin-orbit coupling regime, when characteristic spin-orbit splitting is much less than the level broadening due to scattering. Recent theoretical studies have found the electron spin polarization in weak coupling limit [19, 20, 21, 22], thus confirming the prediction of Refs [10, 11, 12]. However, no scenario of spin generation effect has been considered up to now. A lot of attention was given to the accumulation of spin polarization near the edges of the sample, i.e. SHE [1, 23]. At first glance such an accumulation should take place, similar to the bulk cubic crystals. However, it has been found that the spin-Hall type of accumulation does not exist in the systems [19, 21] whose size is much larger than the spin diffusion length ("cancellation theorem"). Nevertheless such an accumulation can be induced by some "special tricks": cubic in the momentum Dresselhaus spin-orbit terms [21], finite frequency electric field [19], spin relaxation near the edge [24], non-uniform density matrix and application of external magnetic field [22]. Recent experiments revealed the spin Hall type of spin accumulation [25, 26].

Here we apply the KKM theory to the spin Hall current in n-type crystals with reduced symmetry. It is argued that the KKM approach agrees well with the recent theoretical and experimental results and provides their physically transparent interpretation. We show that the spin Hall current *generates* the non-equilibrium spin polarization in the interior of crystals in a way that is drastically different from the previously well-known "equilibrium" polarization during the spin relaxation process. The steady state spin polarization value does not depend on the strength of spin-orbit interaction offering possibility to generate high spin polarization even in the weak spin-orbit coupling case. Finally, we discuss the spin Hall accumulation and show that the "cancellation theorem" does not work in practice, imposing no strong limitations on its experimental observation.

The Hamiltonian of conduction band electron with momentum $\vec{p}$ and effective mass m is $\hat{H} = \frac{p^2}{2m} + \hat{s}_\alpha Q_{\alpha\beta} p_\beta$, where the spin-orbit interaction is characterized by a second rank pseudotensor $Q_{\alpha\beta}$ ($\hat{s}_\alpha$ is the operator of the α-th component of electron spin) [27]. The spin-orbit term can be considered as an interaction of electron spin with the effective magnetic field $\vec{B}_p = \hat{Q}\vec{p}/\mu_B g$ ($\mu_B$>0 is the Bohr magneton, g is the electron g-factor) whose value and direction are determined by those of the electron momentum $\vec{p}$. It brings about the precession of electron spins about $\vec{B}_p$ with frequency $\vec{\Omega}_p = \hat{Q}\vec{p}/\hbar$. For instance, for the Rashba [28] spin-orbit interaction $q\hat{\vec{s}}[\vec{p}\times\vec{n}]$ (asymmetrical quantum wells, strained bulk GaAs, wurtzite-type crystals, etc), one has $Q_{\alpha\beta}=q\varepsilon_{\alpha\beta z}$, the field $\vec{B}_p = q[\vec{p}\times\vec{n}]/\mu_B g$ and the frequency $\vec{\Omega}_p = q[\vec{p}\times\vec{n}]/\hbar$. Here q is spin-orbit constant having the dimensionality of velocity, $\varepsilon_{\alpha\beta\gamma}$ is Levy-Civita symbol, unit vector $\vec{n}$ is parallel to z-axis.

Within the KKM model the spin density $\vec{S}(\vec{r})$ is determined by the semiclassical continuity equation

$$\frac{\partial S_\beta}{\partial t} + \frac{\partial J_{\alpha\beta}}{\partial x_\alpha} = \frac{m}{\hbar}\varepsilon_{\beta j\gamma} Q_{ji} J_{i\gamma} + \left(\frac{\partial S_\beta}{\partial t}\right)_{others} \quad (1)$$

Here $J_{\alpha\beta}$ gives the α-component of the flow of particles with spin polarization along axis β (α,β=x,y,z). It is given by the expectation value of the spin current operator

$$\hat{J}_{\alpha\beta} = \left(\hat{V}_\alpha \hat{s}_\beta + \hat{s}_\beta \hat{V}_\alpha\right)/2, \text{ where the velocity operator } \hat{V}_\alpha = \partial\hat{H}/\partial p_\alpha = p_\alpha/m + Q_{\beta\alpha}\hat{s}_\beta \quad (2)$$

The left-hand side of Eq.(1) gives the time derivative of spin density and the divergence of spin current. The last term on the right-hand side (RHS) takes into account all possible changes of the spin not related with the linear in $\vec{p}$ spin-orbit coupling.

The first term on RHS of Eq.(1) is due to the linear in $\vec{p}$ spin-orbit interaction. It originates from spin precession about $\vec{B}_p$ field with frequency $\vec{\Omega}_p$ and leads to the generation of non-equilibrium spin density $\vec{S}$ in the presence of spin current [13]:

$$\dot{\hat{s}}_\beta = \frac{i}{\hbar}[\hat{H},\hat{s}_\beta] = \left(\vec{\Omega}_p \times \hat{\vec{s}}\right)_\beta = \frac{m}{\hbar}\varepsilon_{\beta j\gamma} Q_{jm}\hat{J}_{m\gamma} \quad (3)$$

We took into account the linear relation $\vec{\Omega}_p = \hat{Q}\vec{p}/\hbar$ and used Eq.(2) to obtain the last equality. Ensemble averaging of Eq. (3) gives the precessional term (RHS of Eq.1). Recently a semiclassical equation similar to Eq.(1) has been derived [20, 29] and the spin-orbit term has been referred to as the "torque density" [29] or the "source" [20] term. Figure 1a illustrates the physics of this term for the case of asymmetric quantum well with normal $\vec{n}\|z$ for Rashba spin-orbit interaction. Suppose a spin current $J_{yz}$ of z-th spin component flows in y direction in the ensemble of initially unpolarized electrons: one half of electrons possess spin up and momentum $\vec{p}$, while another half has spin down and momentum $-\vec{p}$. The spin precession at $\vec{\Omega}_p = q[\vec{p}\times\vec{n}]/\hbar$ and $\vec{\Omega}_{-p} = -\vec{\Omega}_p$ frequencies generates spin density at a rate $\dot{\vec{S}}_p = \dot{\vec{S}}_{-p}$.

The KKM continuity equation (1) imposes certain limitations on the spin current that can flow in low symmetry semiconductor. For example, in steady state regime, in the spatially uniform case and in the absence of other processes not related with linear $\vec{p}$ terms, the Eq. (1) reduces to the precessional term

$$\frac{m}{\hbar}\varepsilon_{\beta j\gamma}Q_{ji}J_{i\gamma}=0 \tag{4}$$

It expresses the "cancellation theorem" [30] because one of its solutions is trivial, $J_{\alpha\beta}=0$ [31]. Note that it is the *total* spin current that enters into Eq. (4). There is many sources contributing to it: (i) skew scattering in external electric field that gives the Dyakonov-Perel spin current [1], (ii) the linear in $\vec{p}$ conduction band spin splitting in the presence of non-equilibrium spin density $\vec{S}$ that induces the KKM spin current [13], (iii) the linear in $\vec{p}$ conduction band spin splitting in the presence of external electric field leading to the spin current [19, 21], sometimes called "intrinsic", etc. The resultant spin current should obey Eq. (4) expressing their mutual compensation [32]. The cancellation is absent in the bulk unstrained GaAs-type crystals where $Q_{\alpha\beta}$=0 and Eq.(4) turns into identity.

Equation (1) shows that it is the conventional spin current $J_{\alpha\beta}$ that should be calculated to obtain the spin density time-space profile [33]. To present the expression for spin current we restrict ourselves by a weak spin-orbit coupling regime. Then the spin current components can be expanded in powers of spin-orbit interaction. In the non-degenerated case the components $J_{\alpha\beta}$ are determined up to the first order in the spin-orbit coupling by the unified DP [1] and KKM [13] equations

$$J_{\alpha\beta}=-bE_\alpha S_\beta - D\frac{\partial S_\beta}{\partial x_\alpha}+\beta N\varepsilon_{\alpha\beta\gamma}E_\gamma + \frac{mD}{\hbar}\varepsilon_{\beta j\gamma}Q_{j\alpha}S_\gamma \tag{5}$$

The first two terms in Eq. (5) take place without spin-orbit interaction and describe the drift of spin density $\vec{S}$ of electrons (with mobility b and concentration N) in an external electric field $\vec{E}$, and spin diffusion with diffusion coefficient D. The last two terms originate from spin-orbit interaction. The third term gives the Dyakonov-Perel spin current (the spin Hall current) resulting from the asymmetric scattering [1, 34]. It exists even in the systems of spherical symmetry and is characterized by the parameter β having the dimensionality of mobility. In contrast to this, the last, KKM, term in Eq.(5) appears only in crystals with a linear in the momentum splitting of the conduction band. It describes the spin current emerging in the presence of non-equilibrium electron spin density $\vec{S}$. Indeed, the ensemble averaged $\langle ... \rangle$ spin current generation rate is [13]

$$\left\langle \dot{\hat{J}}_{\alpha\beta} \right\rangle = \left\langle \frac{i}{\hbar}\left[\hat{H}, \hat{J}_{\alpha\beta}\right] \right\rangle = \left\langle \varepsilon_{\beta i\gamma} \hat{s}_\gamma \frac{p_\alpha Q_{ij} p_j}{m\hbar} \right\rangle = \varepsilon_{\beta i\gamma} S_\gamma \frac{Q_{i\alpha}\langle p_\alpha^2 \rangle}{m\hbar} \quad (6)$$

where $\langle p_\alpha^2 \rangle$ is the mean value of square of momentum of electrons. The generation of spin current is balanced by the fast relaxation with momentum scattering time $\tau_p$. Thus the steady-state value of the KKM spin current is given by the last term in Eq. (5) with the diffusion coefficient $D = \langle p_\alpha^2 \rangle \tau_p / m^2$ (isotropic within this model). Figure 1b illustrates the physics of this term for the Rashba spin-orbit interaction. Let all electrons have spins up. Electrons with oppositely directed momenta $\vec{p}$ and $-\vec{p}$ acquire oppositely directed spin components $\Delta\vec{S}_p$ and $\Delta\vec{S}_{-p} = -\Delta\vec{S}_p$ as a result of spin precession in effective magnetic field with frequencies $\vec{\Omega}_p$ and $\vec{\Omega}_{-p}$. Such a correlation between spin and momentum implies a non-zero current of the y-component of spin in y direction, i.e. $J_{yy} \propto \langle p_y \Delta S_y \rangle \propto S_z$.

The KKM spin current plays very important role for the Eq. (4) to be fulfilled. It may look surprising that all contributions to $J_{\alpha\beta}$ from different physical sources are able to compensate each other to satisfy Eq.(4). The KKM spin current is responsible for it. To do this the electron spin density $\vec{S}$ should be developed, thus generating the KKM spin current. The density $\vec{S}$ is adjusted self-consistently for the KKM current to compensate other currents and satisfy Eq.(4) [35]. Figuratively speaking, the initially flowing spin currents (excepting for the equilibrium ones [32]) are converted *entirely* into the electron spin polarization. In our case the Dyakonov-Perel spin current [1] is compensated by the KKM spin current thus generating the uniform non-equilibrium spin density. Indeed, substituting Eq. (5) into Eq. (1), we have

$$\frac{\partial \vec{S}}{\partial t} = \dot{\vec{S}}_g - \hat{\Gamma}\vec{S} + \vec{\Omega}_{eff} \times \vec{S} \quad (7)$$

The second term in Eq.(7) gives precisely the Dyakonov-Kachorovsky [6] spin relaxation rate with relaxation tensor $\Gamma_{\alpha\beta} = \Gamma_{\beta\alpha} = Dm^2 \left[ Sp(\hat{Q}\hat{Q}^T)\delta_{\alpha\beta} - (\hat{Q}\hat{Q}^T)_{\alpha\beta} \right]/\hbar^2$. The third term describes the precession of $\vec{S}$ in effective magnetic field $\vec{B}_{eff} = \hat{Q}\vec{p}_{dr}/\mu_B g$ with Larmor frequency $\vec{\Omega}_{eff} = \hat{Q}\vec{p}_{dr}/\hbar$ [8] ($\vec{p}_{dr} = -mb\vec{E}$ is the drift momentum of electron ensemble). In contrast to these, the first term of Eq. (7) represents the uniform generation of non-equilibrium spin by the spin Hall current at a rate

$$\dot{\vec{S}}_g = -Nm\beta\hat{Q}^T\vec{E}/\hbar \qquad (8)$$

where $\hat{Q}^T$ is a transposition of matrix $\hat{Q}$. Deducing Eq. (8) we took into account that as a rule $Sp\hat{Q} = 0$ (for Rashba interaction diagonal elements are zero, whereas for symmetric GaAs-type quantum well grown along [001] non-zero components are $Q_{xx} = -Q_{yy}$). We shall illustrate the physical meaning of Eq. (8) for Rashba interaction in asymmetrical quantum well. In this case the DP spin current [1] generates spin density at a rate $\dot{\vec{S}}_g = -N\beta\vec{\Omega}_{eff}/b = Nm q\beta(\vec{E}\times\vec{n})/\hbar$ antiparallel to vector $\vec{\Omega}_{eff}$ and spin precession does not affect $\vec{S}$ (the last term in Eq.(7) is absent). Electric field $\vec{E}\|x$ brings about the drift of electrons (Fig.2). Electrons with opposite spins deflect perpendicular to $\vec{E}$ in opposite directions due to SHE [1]. Two typical trajectories are shown. Spins of electrons at every trajectory rotate about $\vec{\Omega}_p$ ($\vec{\Omega}_{p'}$) direction leading to the generation of spin with a rate $\dot{\vec{S}}_p$ ($\dot{\vec{S}}_{p'}$). Averaging over trajectories gives the mean frequency $\vec{\Omega}_{eff}$ and the mean spin generation rate $\dot{\vec{S}}_g \uparrow\downarrow \vec{\Omega}_{eff}$. The steady-state spin density $\vec{S} = \dot{\vec{S}}_g \tau_s = -N\beta\vec{\Omega}_{eff}\tau_s/b$ accumulated in the sample is the larger, the longer DK spin relaxation time $\tau_s = \Gamma_{xx}^{-1} = \Gamma_{yy}^{-1} = \hbar^2/Dm^2q^2$. Thereby this effect is inherently different from the "equilibrium" spin orientation by electric current [10, 11, 12] due to spin relaxation in $\vec{B}_{eff}$ field. Moreover, the Eq. (5) does not take into account the corresponding ("intrinsic") spin current originating from the linear in $\vec{p}$ conduction band spin splitting in the presence of an external electric field [34, 35]: it is only *quadratic* (unlike the Dyakonov-Perel spin current) on spin-orbit interaction in weak spin-orbit coupling regime [19, 21]. Thus to include the "equilibrium" spin orientation effect [10, 11, 12] into Eq. (7) we have to continue the expansion of $J_{\alpha\beta}$ up to the second order in the spin-orbit interaction.

The degree $\rho$ of steady-state spin polarization

$$\rho = \frac{S}{N} = \frac{\beta}{b}\Omega_{eff}\tau_s = \frac{\hbar}{mD}\frac{\beta E}{q} \qquad (9)$$

does not contain the spin-orbit constant because both $\beta$ and $q$ are linear in spin-orbit interaction. This enables one to increase the value of $\rho$ by choosing an appropriate semiconductor with large $\beta/q$ ratio. The spin-dependent scattering mobility $\beta$ results from the spin-orbit interaction $\lambda\hat{\vec{s}}(\vec{p}\times\nabla V)/2\hbar$ with the potential energy $V(\vec{r})$ being due to impurities [1, 2]. The characteristic spin-orbit parameter $\lambda$ and Bohr radius $a_B$ enable rough estimation of $\beta \sim 2\lambda b/a_B^2 \approx 10^{-3} b$ ($\lambda = 5.3 \overset{\circ}{A}{}^2$, $a_B = 100 \overset{\circ}{A}$ for bulk GaAs) [2]. The simplest way to obtain the $\beta/q$ ratio explicitly is to estimate parameter $q$ for asymmetrical quantum well, where the spin-orbit interaction has a similar form. The only exception is to replace $\nabla V$ by the gradient of the QW potential profile $\langle\nabla U\rangle$ averaged over the wave function of the lowest QW level. In this case parameter $q = \lambda\langle\nabla U\rangle/2\hbar = \lambda eF/2\hbar$ where $F \equiv \langle\nabla U\rangle/e$ is the effective electric field acting on quantized electron from the walls of asymmetric QW (F=0 for symmetric QW). Thus Eq. (9) reads

$$\rho = \frac{\hbar}{mD}\frac{\beta E}{q} \sim \frac{4\hbar^2}{ma_B^2}\frac{bE}{eDF} = \frac{8E_B}{T}\frac{E}{F} \qquad (10)$$

where the Bohr energy $E_B = \hbar^2/2ma_B^2$ ($E_B \approx 5$ meV for bulk GaAs) and we used the Einstein relation $D = bT/e$ to get the last equality. One can see that both spin-orbit parameter $\lambda$ and momentum relaxation time $\tau_p$ are canceled out. The latter takes place because $\beta$ is proportional to mobility $b$, whereas DK spin relaxation time is inversely proportional to $b$. Cancellation, however, is not exact due to the rough estimation of parameter β. Nevertheless, the mobility dependence of ρ may be rather smooth. Estimation for realistic parameters E=100 V/cm, F=$10^5$ V/cm with the use of Eq. (10) gives $\rho \approx 2 \cdot 10^{-3}$ at room temperature for GaAs-type well. As to spin-orbit constant cancellation is concerned, it is robust while the DK spin relaxation dominates. Including of other spin relaxation channel competing to the DK shortens the spin relaxation time. It decreases the result of Eq. (10) when the spin-orbit interaction becomes rather small. Search of optimal conditions and low symmetry semiconductors for the highest electron polarization needs further study and out of scope of this paper.

Substituting the steady-state spin density $\vec{S} = -N\beta\vec{\Omega}_{eff}\tau_s/b$ into Eq. (5) for the spin current we obtain that the last two terms compensate each other and in the linear in $\vec{E}$ approximation spin current is zero in agreement with "cancellation theorem" [36].

Electrically induced non-equilibrium spin polarization can be detected experimentally through its Larmor rotation in an external magnetic field. In steady state regime it leads to the Hanle effect. In non-stationary case spin polarization will oscillate in time. Generation of the uniformly distributed non-equilibrium spin with $\rho \sim 10^{-3} - 10^{-4}$ and its Larmor rotation in an external magnetic field were observed in the Ref. [25, 18] in agreement with this model.

Finally, we can use the KKM approach to discuss the conditions for the spin accumulation near the sample edge, i.e. the appearance of spin polarization different from the bulk value. It follows from Eq.(1) that the *non-uniform* spin polarization $\vec{S}(\vec{r})$ develops provided that the spin current $J_{\alpha\beta}(\vec{r})$ spatially varies. Usually one imposes an "open" boundary condition: the spin current component perpendicular to the boundary is zero [1]. In the conditions of Eq.(4) the spin current is the same ($J_{\alpha\beta} = 0$) both in the bulk and near the boundary leading to the absence of spin accumulation. This is a manifestation of the "cancellation" theorem. In reality, however, the spin currents in the bulk and near the edge are different making the spin accumulation possible. One possibility is that the Eq. (4) is violated, i.e. $J_{\alpha\beta} \neq 0$ in the bulk, as we discussed above. Alternatively, zero boundary condition is violated [22, 37]. For example, an additional spin relaxation may exist at the edge. It implies the non-zero spin current at the boundary, whereas it is still zero in the bulk. In both alternative cases the non-uniform spin current generates the non-equilibrium polarization near the edge. In non-uniform case there is additional contribution into spin density Eq. (7): $b(\vec{E}\nabla)\vec{S} + D\nabla^2\vec{S} - \frac{2mD}{\hbar}(\hat{Q}\nabla \times \vec{S})$. The first two terms describe the usual drift-diffusion motion of average spin density. The last term has spin-orbit origin. It was derived first in [9, 13] and later in Refs. [15, 20, 21]. One of its meanings is the spin precession in doubled effective field due to diffusion motion. It is responsible for the oscillations of non-uniform $\vec{S}$ in space [20, 21, 37, 38] even in the absence of external magnetic field in agreement with experiment [15, 25]. An example of SHE oscillations for the case of the Dyakonov-Perel spin current is considered in Ref. [39].

In conclusion, we applied the KKM approach to discuss the relationship between non-equilibrium spin density and spin Hall current due to asymmetry in scattering. It is shown that the spin Hall current *generates* the non-equilibrium spin polarization in the interior of crystals in a way that is drastically different from the previously well-known "equilibrium" polarization during the spin relaxation process. The steady state spin polarization value does not depend on the strength of spin-orbit interaction offering possibility to generate relatively high spin polarization even in the weak spin-orbit coupling case.

Author is grateful to E.L. Ivchenko, K.V. Kavokin and I.A. Merkulov for fruitful discussions. The paper is supported by CRDF, RSSF, RFBR grants, and the Department of Physical Sciences and the Presidium of the Russian Academy of Sciences.

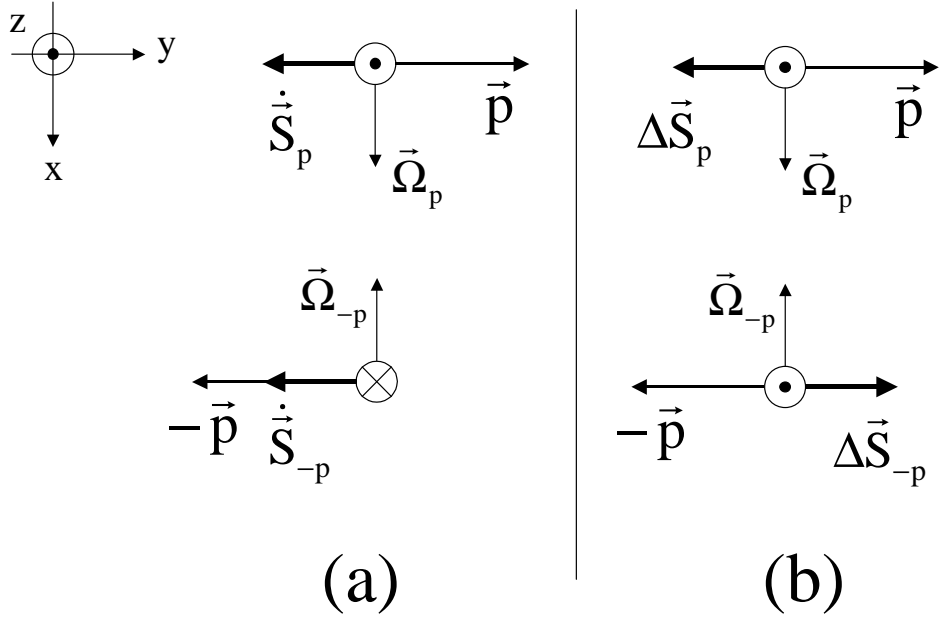

(a) Illustration of the physics of the precessional term. Spin current generates spin polarization of electrons. Suppose a spin current $J_{yz}$ of z-th spin component flows in **y** direction in the ensemble of initially unpolarized electrons. It means that one half of electrons has spin up ($\bullet$) and momentum $\vec{p}$, whereas another half has spin down ($\otimes$) and momentum $-\vec{p}$. The spin precession with $\vec{\Omega}_p$ and $\vec{\Omega}_{-p} = -\vec{\Omega}_p$ frequencies generates spin at a rate $\dot{\vec{S}}_p = \dot{\vec{S}}_{-p}$ opposite to **y**.

(b) Origin of the KKM spin current. Spin polarized electron ensemble (all spins look up) creates spin current. Electrons with oppositely directed momenta $\vec{p}$ and $-\vec{p}$ acquire oppositely directed spin components $\Delta\vec{S}_p$ and $\Delta\vec{S}_{-p}$ as a result of spin precession with frequencies $\vec{\Omega}_p$ and $\vec{\Omega}_{-p}$. Such a spin-momentum correlation implies a non-zero $J_{yy} \propto \langle p_y \Delta S_y \rangle$

Korenev, Figure 2

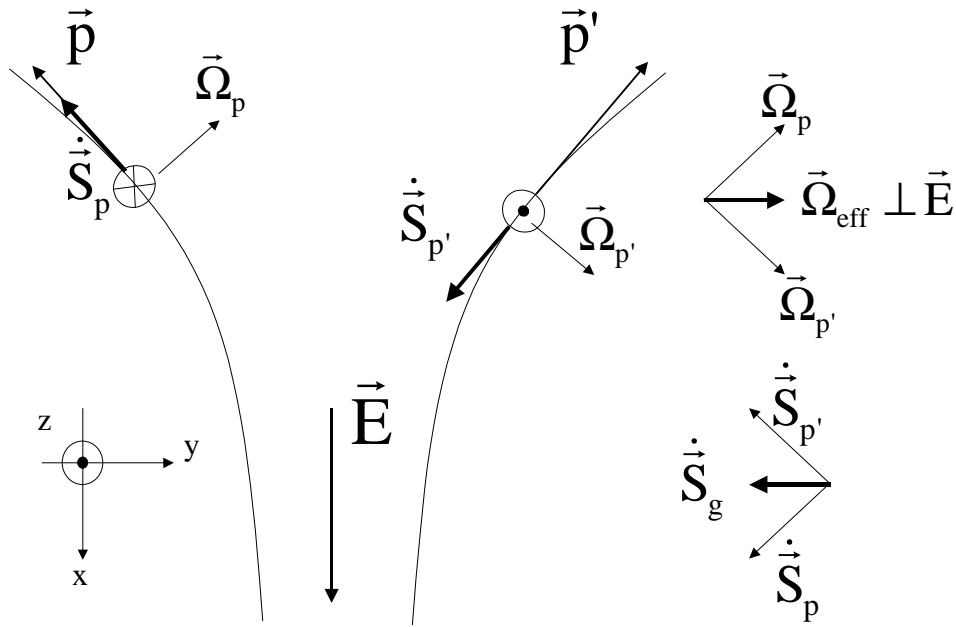

Spin Hall current generates electron spin. Electric field $\vec{E} \| x$ brings about the drift of electrons opposite to it. Electrons with opposite spins deflect in opposite directions perpendicular to the field [1]. Two typical trajectories are shown. Spins of electrons at every trajectory rotate about $\vec{\Omega}_p$ ($\vec{\Omega}_{p'}$) direction leading to the generation of spin at a rate $\dot{\vec{S}}_p$ ($\dot{\vec{S}}_{p'}$). Averaging over the trajectories gives the mean spin *generation rate* $\dot{\vec{S}}_g$ antiparallel to vector $\vec{\Omega}_{eff}$.